\documentclass[aps,amsmath]{revtex4}
\usepackage{amssymb}
\usepackage{graphicx}
\usepackage{bm}
\usepackage[latin1]{inputenc}
\usepackage[portuguese]{babel}
\usepackage{graphicx}
\usepackage{amssymb}
\usepackage{amsthm}
\usepackage{dcolumn}
\usepackage{geometry}
\begin{document}
\markboth{F.S. Borcsik and S.D. Campos}{Odderon and Pomeron as Fractal Dimension in $pp$ and $\bar{p}p$ Total Cross Sections}

%
%




\title{Odderon and Pomeron as Fractal Dimension in $pp$ and $\bar{p}p$ Total Cross Sections}
\author{F. S. Borcsik and S. D. Campos\footnote{E-mail: sergiodc@ufscar.br}}
\address{Departamento de F\'isica, Qu\'imica e Matem\'atica, Universidade Federal de S\~ao Carlos, CEP 18052780, Sorocaba, SP, Brazil.}





\begin{abstract}
In this paper one presents a naive parametrization to $pp$ and $\bar{p}p$ total cross sections. The main goal of this parametrization is to study the possible fractal structure present in the total cross section. The result of the fitting procedure shows two different fractal dimensions: a negative (low-energies) and a positive (high-energies). The negative fractal dimension represents the emptiness of the total cross section structure and the positive represents the filling up process with the energy increase. Hence, the total cross section presents a multifractal behavior. At low-energies, the odderon exchange may be associated with the negative fractal dimension and at high-energies, the pomeron may be associated with the positive fractal dimension. Therefore, the exchange of odderons and pomerons may be viewed as a transition from a less well-defined to a more well-defined internal structure, depending on the energy.

\end{abstract}

\maketitle


\section{Introduction}
Since the Mandelbrot's seminal papers \cite{mandelbrot_1,mandelbrot_20,mandelbrot_30,mandelbrot_40,mandelbrot_50} in the beginning of 1960s, the mathematical understanding of fractal geometry was developed and applied in physics, mathematics, statistics, geophysics and biology (among other sciences). The advantage claimed by fractal geometry over Euclidean rests on the possibility to describe irregular or fragmented shapes present in nature providing scaling properties, allowing different points of view to a given problem.   

A fractal may be viewed as a conceptual object possessing geometrical properties, classified as uniform or non-uniform. The first one is also known as self-similar, and the second is the so-called multifractal. The mathematical notion of self-similarity defines structures or sets which have only one scaling factor (one scaling law). On the other hand, multifractality is related to structures or sets (datasets) possessing two or more scaling factors. There are in the literature several ways to define and compute the fractal dimension of a given structure and the most common are the Hausdorff-Besicovitch dimension \cite{hausdorff}, and as defined by Mandelbrot \cite{mandelbrot_1}. However, there are other mathematical methods as box-counting, hyperbolic functions, Fourier transforms, wavelets, etc. \cite{li}.

Probably, the first paper relating a scattering process and fractals is due to Berry introducing the term diffractals to describe waves diffracted by fractals \cite{berry}. Thenceforth, several papers regarding this subject were proposed (see \cite{lidar} and references therein). The notion of fractality was also used to study intermittency in multiparticle production \cite{dremin_1,hakioglu,angelis,turko,gustafson,bialas,bialas_10,peschanski,antoniou_1,antoniou_10,antoniou_20,antoniou_30,antoniou_2} (and references therein). Properties of the proton structure at low Bjorken $x$ region are also described using the pattern of self-similar fractals \cite{lastovicka}. Soft hadronic processes and fractals were considered in \cite{carruthers,dremin_2,dremin_200}. The fractal pomeron as a mechanism related to the onset of a second-order phase transition in the Feynman-Wilson fluid in the rapidity space is presented in \cite{antoniou_1,antoniou_10,antoniou_20,antoniou_30}. Details about hadronic processes and self-similarity may be viewed in \cite{kittel}.

In the present paper one analyzes proton-proton ($pp$) and antiproton-proton ($\bar{p}p$) total cross sections ($\sigma_{tot}(s)$) by means of a naive parametrization based on the Froissart-Martin bound \cite{froissart_martin,froissart_martin_2,froissart_martin_3}. The main goal of this parametrization is to define and measure the fractal behavior of $\sigma_{tot}(s)$ and not to obtain the \textit{best description} of the datasets. In this way, independent fits to each available dataset were performed to $pp$ and $\bar{p}p$ total cross section. The aim of this procedure is to prevent that $pp$ data at low-energies counterweight the absence of $\bar{p}p$ data at same energy range as well as $\bar{p}p$ at high-energies counterweight the lack of $pp$ experimental data. 

In order to avoid any misunderstanding, it is important to stress that fractal dimension is \textit{not} the topological dimension of a set. The topological dimension is always an integer number and the fractal dimension may assume non integers and negative values \cite{mandelbrot_2,mandelbrot_300,mandelbrot_200}, complex \cite{pietronero}, and fuzzy values \cite{feng}. Keeping it in mind, it is shown that odderons and pomerons may be identified to different fractal dimensions obtained from the fitting procedures to the total cross sections. It may allow a new point of view to the filling-up process of the total cross section. 

The paper is organized as follows. In the section \ref{fd} one introduces the mathematical definition of fractal dimension. In the section \ref{tcs} basic assumptions on total cross section are presented. Fitting results to the total cross sections are shown in the section \ref{fit}.  
Finally, section \ref{fr} presents discussions upon the results and critical remarks.

\section{\label{fd}Fractal Dimension: Basic Assumptions}
Consider a closed bound set $X\subset \mathrm{R}^n$. If $X$ is a manifold, then its Euclidean topological dimension is an integer number. For general sets, the dimension of $X$ is not necessarily an integer number as the Julia set, for instance \cite{milnor}. 
It is possible to define the topological dimension of a set $X$ by induction: $X$ has zero-dimension if for every point $x\in X$ every sufficiently small ball about $x$ has boundary not intersecting $X$ ($dim_T(X)=0$). The dimension of $X$ is $d$ if for every point $x\in X$ every sufficiently small ball about $x$ has no boundary intersecting $X$ in a set of dimension $d-1$. This definition coincides with the usual notion of dimension for manifolds and of course, it assumes only integer values. See, for instance, Coornaert and references therein for a more detailed study \cite{coornaert}.

In general, one uses here a more handleable method to measure a fractal dimension. One defines a dimension of a set using the box-counting method: given $\epsilon>0$, let $N(\epsilon)$ be the smallest number of $\epsilon$-balls needed to cover $X$. The box dimension $dim_B(X)$ can be written as
\begin{eqnarray}
\label{bd} dim_B(X)=\lim \limits_{\epsilon\rightarrow 0} \mathrm{sup}\frac{\log N(\epsilon)}{\log (1/\epsilon)}.
\end{eqnarray}  

The above definition coincides for manifolds and $\epsilon$ represents the magnification scale of the structure. On the other hand, the countable sets of points have half-dimension using the box counting. This problem can be solved using the Hausdorff-Besicovitch definition: given $X$ one considers a cover $U=\{U_i\}_i$ for $X$ by open sets. For $\delta>0$ one define $H_{\epsilon}^\delta(X)=\mathrm{inf}_U\{\sum_i \mathrm{diam}(U_i)^\delta\}$ where the infimum is taken over all open covers $U=\{U_i\}$ such that $\mathrm{diam}(U_i)\leq\epsilon$. Defining $H^\delta(X)=\lim \limits_{\epsilon\rightarrow 0}H_{\epsilon}^\delta(X)$ one writes
\begin{eqnarray}
\label{hb} dim_{HB}(X)=\mathrm{inf}\{\delta:H^\delta(X)=0\}.
\end{eqnarray}

A fractal can also be defined by using Mandelbrot: a fractal is a set whose Hausdorff-Besicovitch dimension strictly exceeds its topological dimension \cite{mandelbrot_2,mandelbrot_300,mandelbrot_200}. For the sake of simplicity, one introduces the mass-radius method. It consists in covering a structure with discs and then counting the number of points (elements) inside each disc. Mathematically, one writes
\begin{eqnarray}
\label{mr} M_r(A)=\mu \left(\frac{r}{r_0}\right)^D,
\end{eqnarray}

\noindent where $M_r(A)$ is the number of points of $A$ inside the disc of radius $r_0<r$ and $\mu$ is a positive real number. The fractal dimension $D$ is obtained when the following limit exist
\begin{eqnarray}
\label{mr1}D=\lim\limits_{r\rightarrow \infty} \frac{\ln (M_r(A)/\mu)}{\ln(r/r_0)}.
\end{eqnarray}

The meaning of (\ref{mr1}) is obvious: when $r\rightarrow \infty$, the number of discs of fixed-radius $r_0$ used to cover some structure (a line or a map, for instance) increases, revealing more details about the fractal structure of the object. 

The main characteristic of a fractal is its self-reproducing pattern. Self-similar fractals with an appropriate magnification show a multiple reproduction of themselves. In nature, the self-similarity is an approximation that may be used to describe the physical behavior of a given system. 

In the general theory of fractals, a negative dimension represents a measure of the emptiness of the structure, and positive values can be viewed as the self-similarity of the structure \cite{mandelbrot_2,mandelbrot_300,mandelbrot_200}.  

As naive example, consider the classical total cross section of a hard sphere which is one of the most interesting observables of the elastic scattering process. Its geometrical description is simple and provides an intuitive picture of the hadron-hadron elastic scattering \cite{martinmatthiae}. From the geometrical point of view, the description of the elastic scattering by a hard sphere of radius $R$ uses the impact parameter $b=R\sin \psi$ and the scattering angle, $\theta=\pi-2\psi$. Therefore, $b=R\cos(\theta/2)$. In this picture, the total cross section $\sigma$ is written as
\begin{eqnarray}
\label{tcsb}\sigma=\pi R^2.
\end{eqnarray}

Using (\ref{mr1}) one obtains to the classical total cross section a fractal dimension of 2 in agreement with its topological dimension since every (filled) circle is diffeomorphic to all $\mathbb{R}^2$. However, this is a classical geometrical view and does not correspond to the total cross section obtained from a particle-particle or antiparticle-particle elastic scattering. The non-classical elastic scattering at high-energies is far from be understood, resulting in several models used to describe its behavior \cite{blockwhite,dl,dl2,sdc_1}. In the next section one presents a naive parametrization based on the Froissart-Martin bound to describe the total cross section as a toy-model to study its fractal behavior.

\section{\label{tcs}Total Cross Section}
As well-known, total cross section at high-energies is dominated by the Froissart-Martin bound \cite{froissart_martin,froissart_martin_2,froissart_martin_3} written as
\begin{eqnarray}
\label{fm1} \sigma_{tot}(s)\leq \alpha\ln^2(s/s_0),
\end{eqnarray} 

\noindent where $m_\pi$ is the pion mass, $s$ is the square of the center-of-mass energy, $\alpha=\frac{\pi}{m_\pi^2}$ and, usually, $s_0=1.0$ GeV$^2$. This is a bound obtained in the asymptotic regime ($s\rightarrow\infty$). Therefore, one cannot expect its saturation at current energies. The search for a better mathematical description of $\sigma_{tot}(s)$ includes a search for a smaller $\alpha$. As an effort to improve this bound, Martin ``showed'' \cite{martin_2009}
\begin{eqnarray}
\label{fm2} \sigma_{tot}(s)\leq \frac{\alpha}{2}\ln^2(s/s_0).
\end{eqnarray} 

\begin{figure}[ht]
\centering{\includegraphics[scale=0.2]{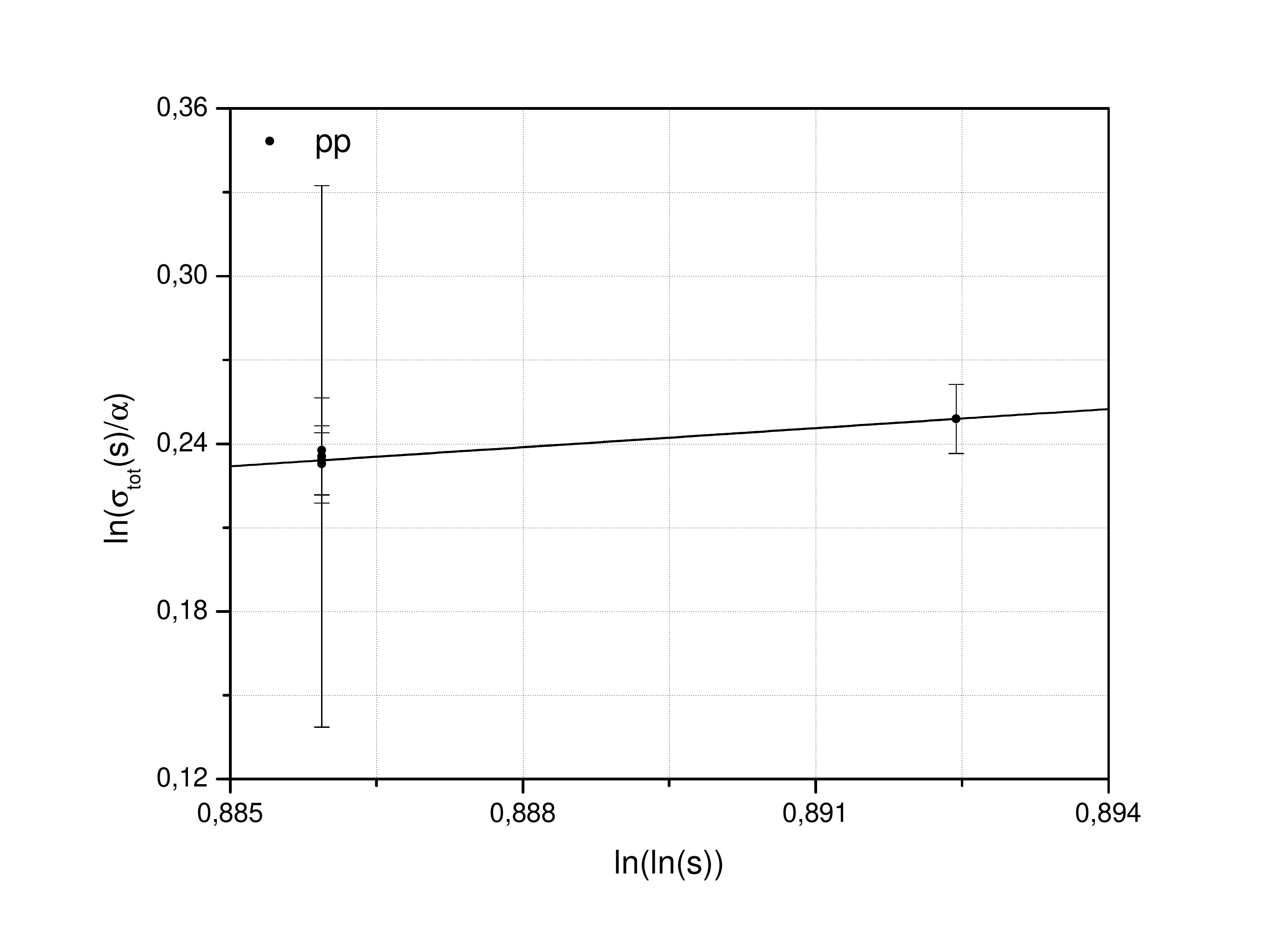}}
\centering{\includegraphics[scale=0.2]{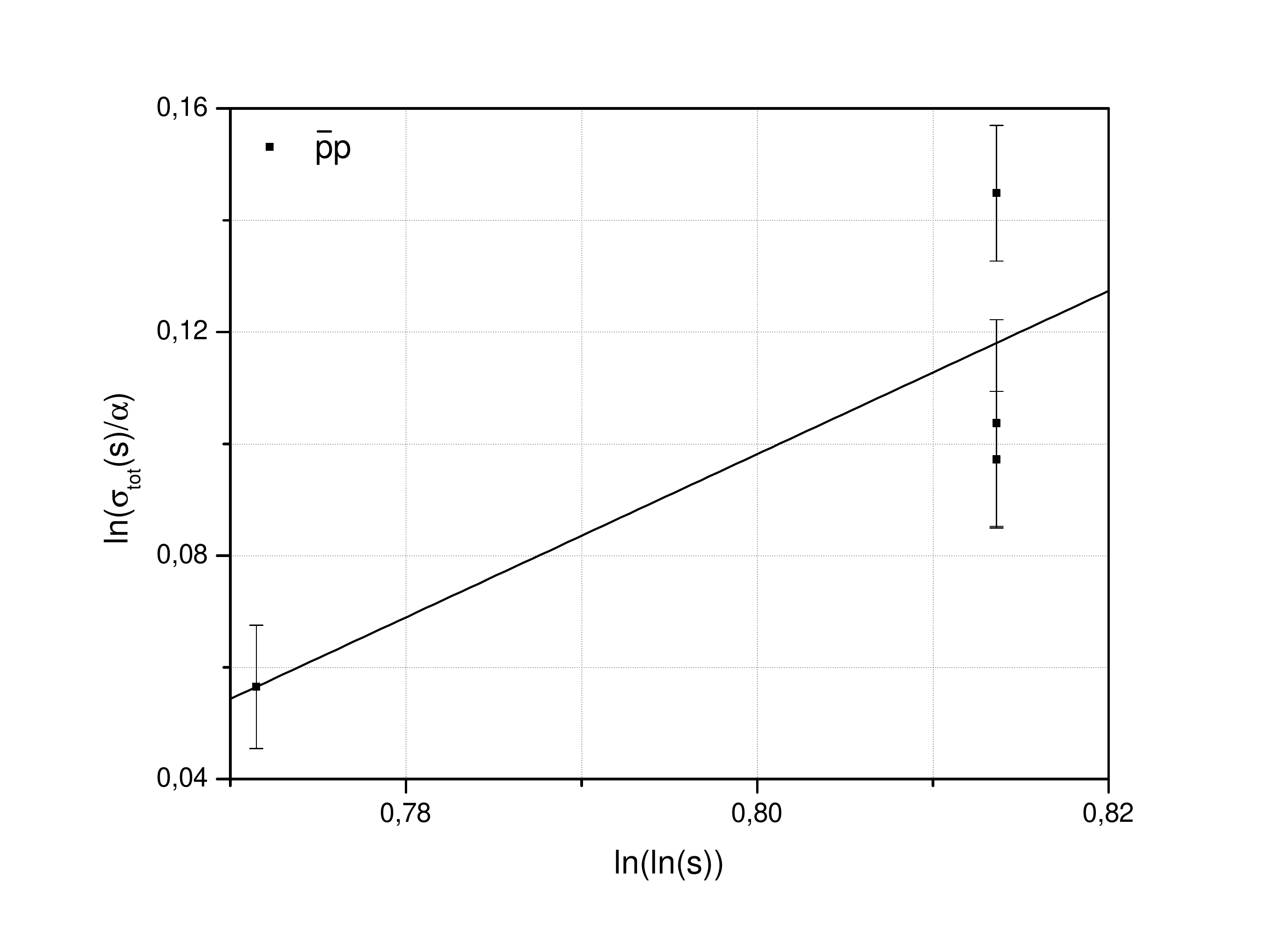}}
\vspace{-0.6cm}
\caption{Plots of $pp$ and $\bar{p}p$ total cross sections above $\sqrt{s}=50.0$ GeV. These data may well be described by a straight-line.}
\label{fig1}
\end{figure}

To our purposes, one uses here only (\ref{fm1}), since $\alpha$ is just a scale factor. Keeping in mind that (\ref{fm1}) is an asymptotic condition, one plots $\ln(\sigma_{tot}(s)/\alpha)$ $versus$ $\ln(\ln(s))$ using only $pp$ and $\bar{p}p$ dataset at very-high-energies. Specifically, one uses the TOTEM data at $\sqrt{s}=7.0$ TeV \cite{totem} and at $\sqrt{s}=8.0$ TeV \cite{totem2}. The data to $pp$ from cosmic-ray were excluded since they are model-dependent \cite{regina}. Figure \ref{fig1} shows both graphics and to fit these data one uses a simple straight-line
\begin{eqnarray}
\nonumber y=A_1x+A_2,\\
\nonumber \bar{y}=\bar{A}_1x+\bar{A}_2,
\end{eqnarray}

\noindent and obtains to $pp$ $A_1=1.46\pm0.66$ and $A_2=-1.07\pm 0.52$ ($\chi^2/dof=4.19$); to $\bar{p}p$ one obtains $\bar{A}_1=2.28\pm 0.29$ and $\bar{A}_2=-1.78\pm0.25$ ($\chi^2/dof=0.02$). These results show that $pp$ is far from the saturation condition. However, $\bar{p}p$ saturates at the present day energies. 

If total cross section saturates, then it is possible to write
\begin{eqnarray}
\label{fb} \sigma_{tot}(s)=\beta\ln^2(s), \hspace{0.3cm} (s\rightarrow\infty)
\end{eqnarray}

\noindent where $\beta>0$ is a free parameter. In a straight analogy with (\ref{mr1}) it can be written

\begin{eqnarray}
\label{fd1} 2=\frac{\ln\left(\frac{\sigma_{tot}(s)}{\beta}\right)}{\ln(\ln(s))}, \hspace{0.3cm} (s\rightarrow\infty)
\end{eqnarray}

\noindent where the natural length scale in the energy space is given in unities of $\ln(s)$ and $\sigma_{tot}(s)/\beta$ is simply the reduced total cross section. It is important to stress that the above result does not represent a topological dimension. As shown below, the fractal dimension associated to $\sigma_{tot}(s)$ will represent the odderon and pomeron contributions in the scattering process at low or high-energies, respectively.   

In order to analyze the problem of saturation of the Froissart-Martin bound one proposes the following parametrizations to the total cross section of $pp$ and $\bar{p}p$ written as
\begin{eqnarray}
\label{mod1}\sigma_{tot}^{pp}(s)=a\ln^b(s/s_0)\\
\label{mod2}\sigma_{tot}^{\bar{p}p}(s)=\bar{a}\ln^{\bar{b}}(s/s_0)
\end{eqnarray}

\noindent where $a(\bar{a})$ and $b(\bar{b})$ are free fit parameters. These parametrizations will be used in three different regimes: a low-energy ($2.5$ GeV$<\sqrt{s}<20.0$ GeV), a high-energy ($20.0$ GeV$\leq$ $\sqrt{s}< 1.0$ TeV) and very-high-energy ($\sqrt{s}\geq 1.0$ TeV). The purpose is to study a possible fractal point of view of the total cross section, connecting $b$ and $\bar{b}$ to the odderon and pomeron pictures at different regimes.  It is important to emphasize that odderons and pomerons are initially proposed as mathematical tools (Regge trajectories, in fact) to explain the total cross section behavior at high energies. However, as noted by Martynov \cite{martynov}, the TOTEM collaboration measured, in fact, the soft pomeron contribution (pomeron and odderon).

\section{\label{fit}Fitting Results}
Fits to the total cross section datasets were performed in two regimes: low and high-energy. Low-energy corresponds to $\sigma_{tot}^{pp}(s)$ $2.5$ GeV $<\sqrt{s}<20.0$ GeV (123 experimental values) and to $\sigma_{tot}^{\bar{p}p}(s)$  $3.5$ GeV $<\sqrt{s}<20.0$ GeV (50 experimental values). High-energy corresponds to $\sigma_{tot}^{pp}(s)$ $\sqrt{s}\geq 20.0$ GeV (47 experimental values) and to $\sigma_{tot}^{\bar{p}p}(s)$ $\sqrt{s}\geq 20.0$ GeV (17 experimental values). Experimental datasets were taken from Particle Data Group archives \cite{pdg}. The recent datasets to $\sigma_{tot}^{pp}(s)$ at $\sqrt{s}=7.0$ TeV \cite{totem} and 8.0 TeV \cite{totem2} were used. No constraints on the fitting parameters were imposed in the fitting procedures. The obvious consequence of this approach that follows is that one may not be able to verify the modified Pomeranchuk theorem, written as
\begin{eqnarray}
\label{pom} \frac{\sigma_{tot}^{\bar{p}p}}{\sigma_{tot}^{pp}}\rightarrow 1, \hspace{0.3cm} (s\rightarrow\infty).
\end{eqnarray}

At finite energies one cannot expect the confirmation of all asymptotic theorems as Pomeranchuk, Froissart-Martin bound, among others, in the context of AQFT. In some situations, an asymptotic condition may occur (saturate), and in others it may not, since there is no value of $s$ to start its applicability. Therefore, the Froissart-Martin bound could be saturated and (\ref{pom}) may not. To impose that all asymptotic conditions must be satisfied by a model would be quite artificial, in fact.

These theoretical results may provide insights to the proposal of novel models. They cannot by itself refuse any model at present day energies. It is not a real problem, since one of the most important models describing the $pp$ and $\bar{p}p$ total cross section possesses a parametrization that violates the Froissart-Martin bound \cite{dl,dl2} and no one can deny its incredible physical content. Therefore, parametrizations to the total cross section may violate some important theorems and bounds from AQFT \textit{at finite energies}. These theorems, however, should be restored via some unknown physical saturation mechanism, for instance.

\begin{figure}
\centering{\includegraphics[scale=0.2]{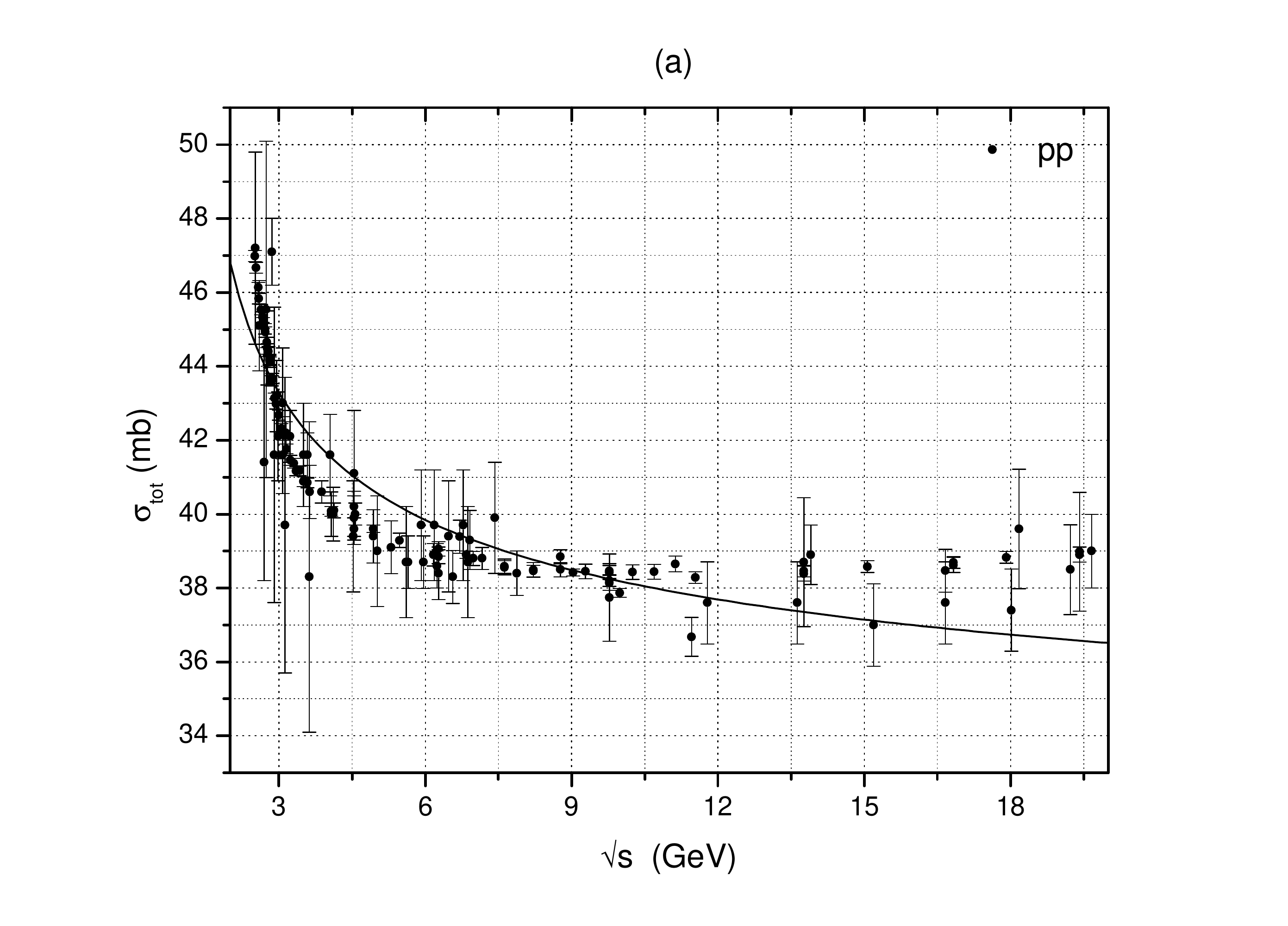}}
\centering{\includegraphics[scale=0.2]{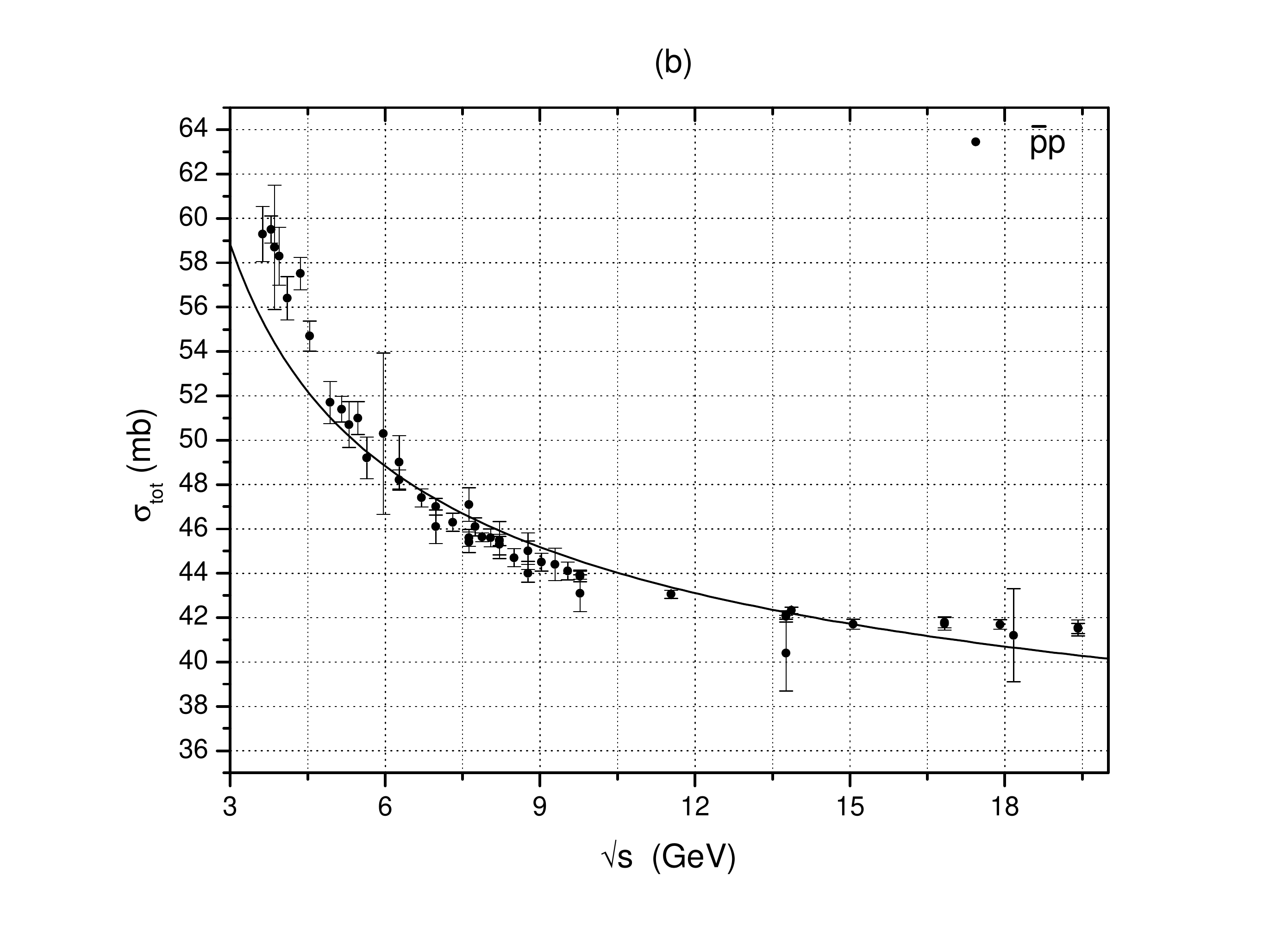}}
\caption{\label{fig2}Results to (a) $pp$ and (b) $\bar{p}p$ total cross sections using parametrization (\ref{mod1}) and (\ref{mod2}) at low-regime.}
\end{figure}

\begin{figure}
\centering{\includegraphics[scale=0.2]{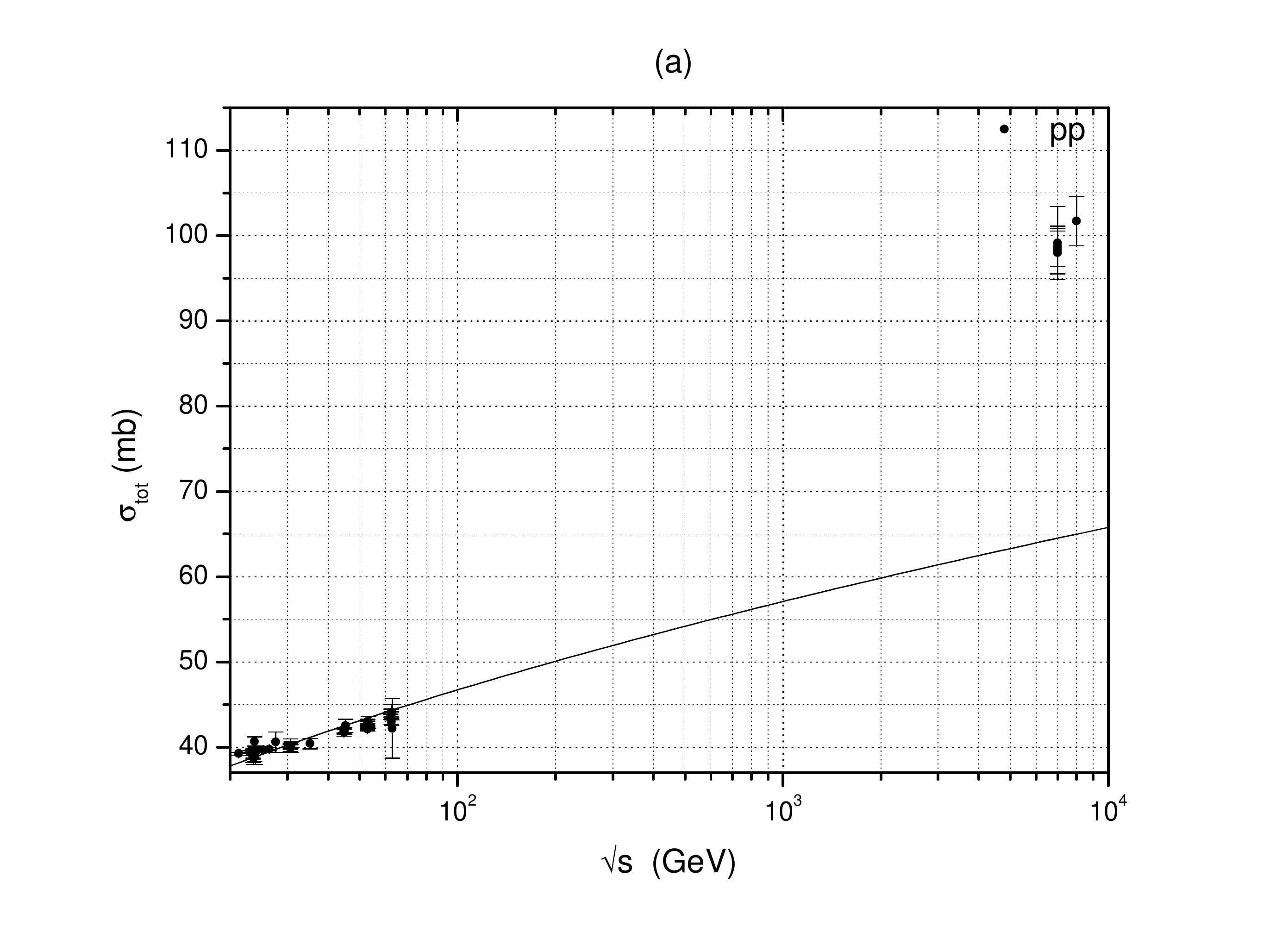}}
\centering{\includegraphics[scale=0.2]{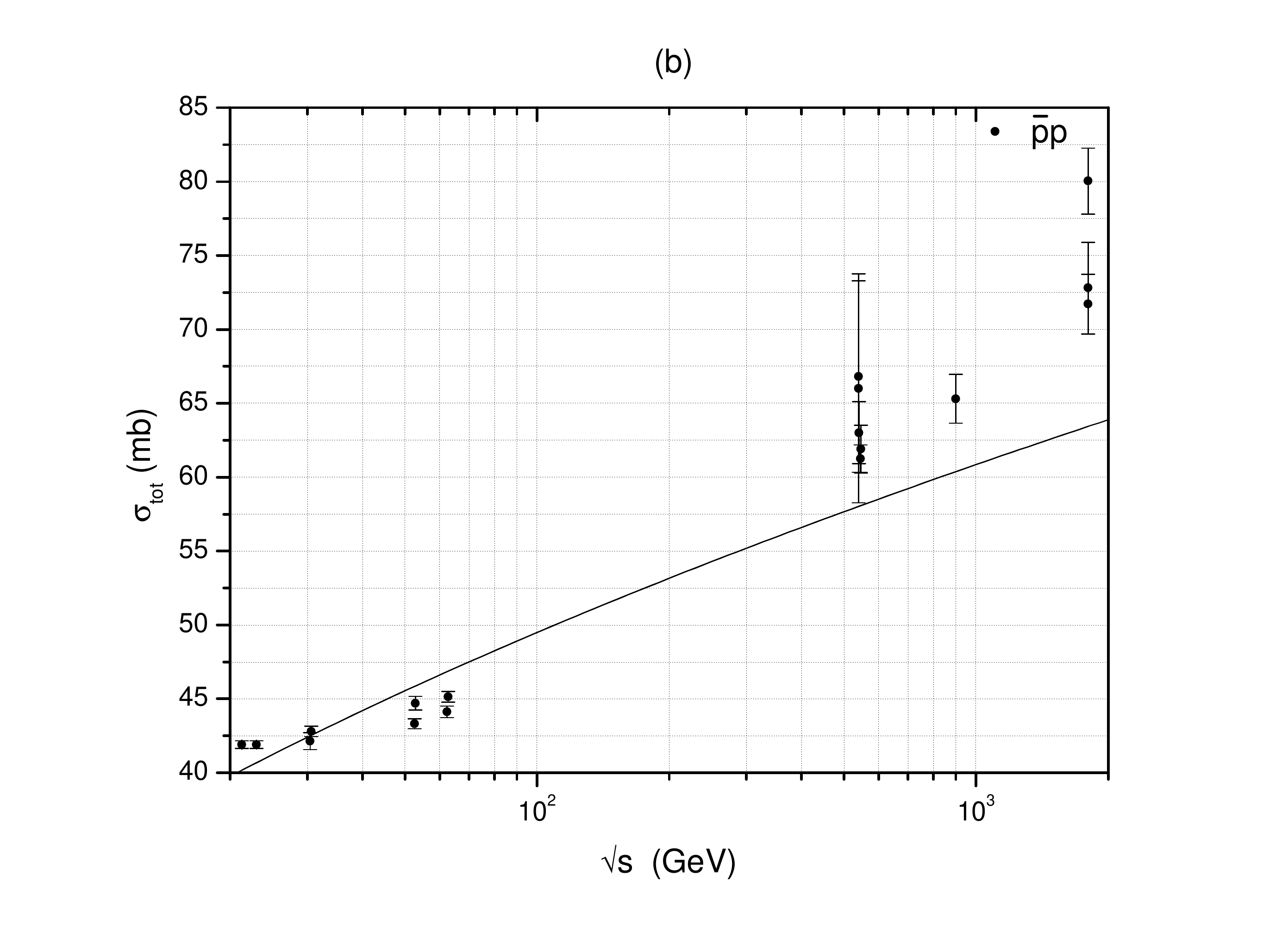}}
\centering{\includegraphics[scale=0.2]{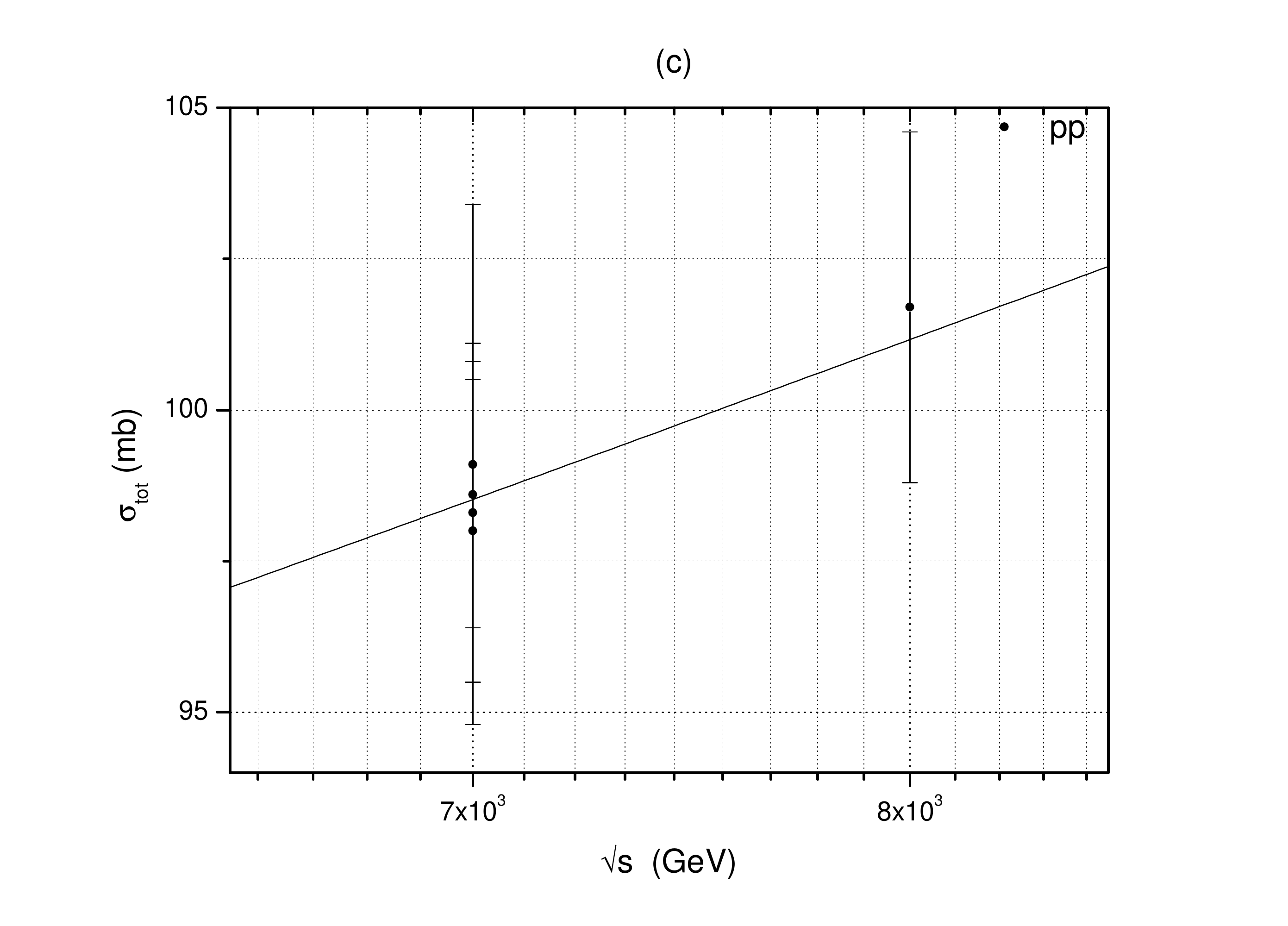}}
\centering{\includegraphics[scale=0.2]{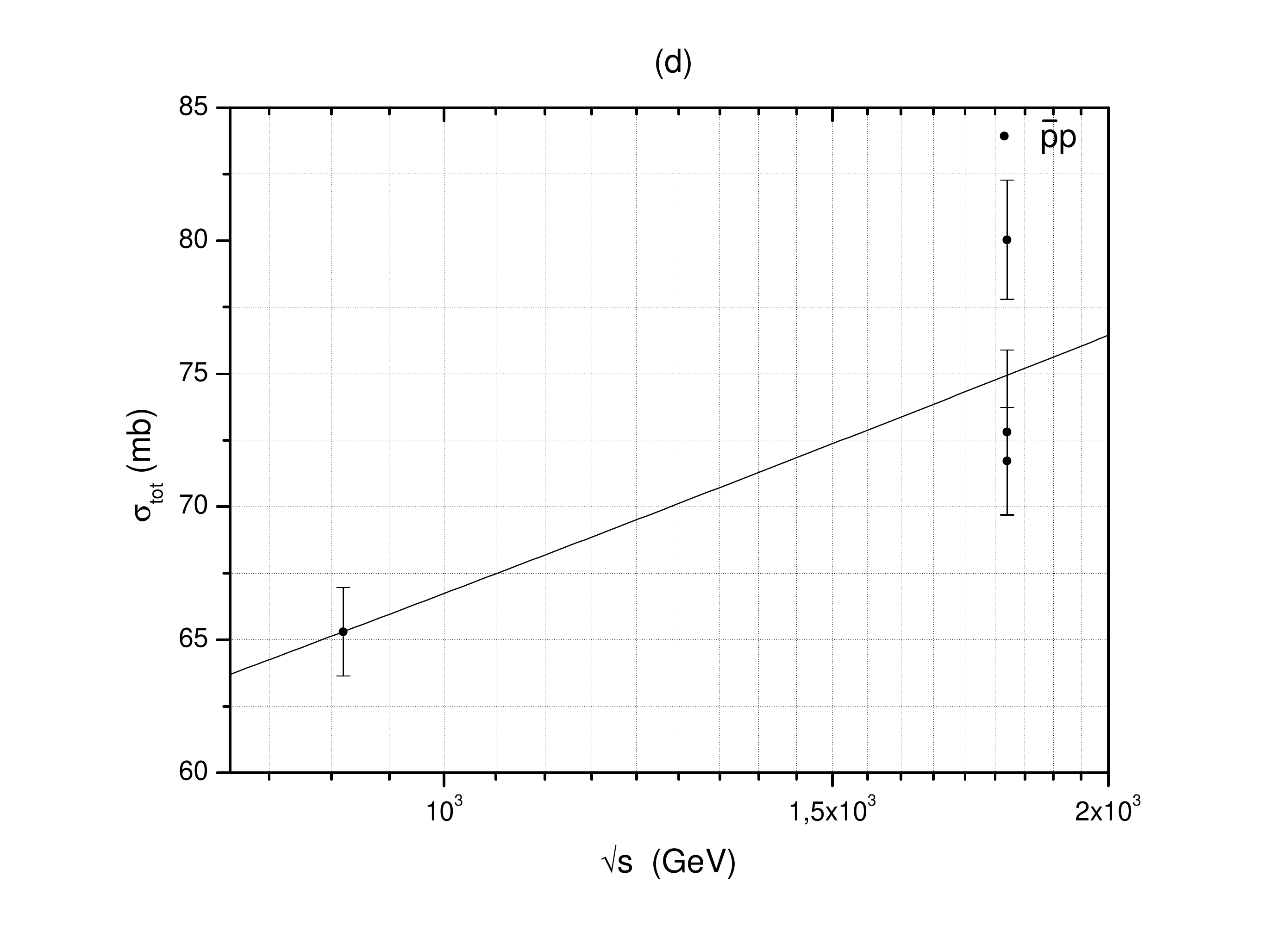}}
\caption{\label{fig3}Results to $pp$ and $\bar{p}p$ total cross sections using parametrization (\ref{mod1}) and (\ref{mod2}).}
\end{figure}

Table \ref{table1} shows the fitting results to $\sigma_{tot}(s)$ at low and high-energies. As commented before, the aim of this work is not to obtain the best fit result since the model proposed is just a way to study the fractal behavior of odderon and pomeron pictures. Figure \ref{fig2} shows the fit results using parametrizations (\ref{mod1}) and (\ref{mod2}) at low-energies. Table \ref{tablesobe} shows the results using only $pp$ and $\bar{p}p$ data at very-high-energies. Figure \ref{fig3} shows the fit results at very-high-energies.

\begin{table*}[ht]
\caption{Fit results to $pp$ and $\bar{p}p$ total cross sections using parametrization (\ref{mod1}) and (\ref{mod2}) at low- and high-energies.}
{\begin{tabular}{cc|cc|cc|cc}
\hline
\multicolumn{4}{c|}{low-energy} & \multicolumn{4}{|c}{high-energy}  \\  
\hline
$pp$              & & $\bar{p}p$                             & &  $pp$                  & &  $\bar{p}p$ &  \\
\hline
$a$     & 49.47$\pm$0.35 &    $\bar{a}$   & 79.37$\pm$2.09   & $a$     & 15.62$\pm$1.37 & $\bar{a}$   & 15.98$\pm$1.98  \\
\hline
$b$     & -0.17$\pm$0.01 &    $\bar{b}$   &-0.38$\pm$0.02    & $b$     & 0.49$\pm$0.04  & $\bar{b}$   & 0.51$\pm$0.06    \\
\hline 
 & $\chi^2/dof=26.6$         & &  $\chi^2/dof= 6.75$          & &  $\chi^2/dof= 23.5$      & & $\chi^2/dof=20.70$        \\
\hline
\end{tabular}\label{table1}}
\end{table*}

\begin{table*}[ht]
\caption{Fit results to $pp$ and $\bar{p}p$ total cross sections using parametrization (\ref{mod1}) and (\ref{mod2}) only at very-high-energies.}
{\begin{tabular}{cc|cc}
\hline
\multicolumn{4}{c}{very-high-energy} \\
\hline
$pp$              &   &$\bar{p}p$          &               \\ 
\hline
$a$     & 0.61$\pm$0.68 &    $\bar{a}$   & 1.61$\pm$2.77   \\
\hline
$b$     & 1.77$\pm$0.38 &    $\bar{b}$   &1.42$\pm$0.64   \\ 
\hline 
 & $\chi^2/dof=0.03$         & &  $\chi^2/dof= 4.1$          \\ 
\hline
\end{tabular}\label{tablesobe}}
\end{table*}

The difference observed in the slopes of $pp$ and $\bar{p}p$ total cross sections at low-energies (see Table \ref{table1}) is attributed, in general, to the odderon exchange as being the leading process (odderon has negative parity). Hence, odderons may be represented as the negative fractal dimension obtained in the fitting procedure. In the fractal theory, the negative dimension represents a measure of the emptiness of some structure \cite{mandelbrot_2,mandelbrot_300,mandelbrot_200}. Hence, odderons may be viewed as a measure of the emptiness present in the $\sigma_{tot}(s)$, i.e., at low energies it represents the absence of a well-defined internal structure of the colliding particles. Therefore, the minimum at the total cross section dataset represents the smallest emptiness value in the colliding structure. In terms of a geometrical picture, the density of the colliding particles increase toward the center, approaching the black disk description.   

On the other hand, the increasing of $pp$ and $\bar{p}p$ total cross sections is a well-known fact and it is expected that $\sigma_{tot}^{pp}(s)\rightarrow\sigma_{tot}^{\bar{p}p}(s)$ as $s\rightarrow\infty$. This result is attributed to the presence of a pomeron exchange (this is the original Pomeranchuk's theorem) since it does not differ particles from antiparticles (pomeron has positive parity). Thus, the positivity of fractal dimension may be associated to a pomeron exchange. If the fractal dimension tends to the Froissart-Martin bound saturation, then the internal structure tends to the maximum of its well-defined state.

As well known, the elastic scattering is the shadow of particle production (inelastic cross-section). If one takes into account the phenomenon of intermittency, then the spectrum of the produced particles in terms of the scaled factorial moments obeys a power-law depending on the phase-space volumes and the number of particles in a given volume \cite{bialas,bialas_10,antoniou_2}. At the same time, this power-law controls the particle density over the momentum space \cite{bialas,bialas_10}. 

Recently, the authors of \cite{antoniou_2} show that a fractal structure emerges from a critical point in configuration space, inducing multidimensional intermittency patterns in momentum space, in accordance with experimental evidence. Following this idea, one may expect that momentum correlations may become important observables for the detection of critical fluctuations \cite{antoniou_1,antoniou_10,antoniou_20,antoniou_30} implying, for instance, that a power-law controlling the fast rise of the $\sigma_{tot}(s)$ when Froissart-Martin saturation occurs. 

If one considers this turning point (the zero of the fractal dimension) as a symmetry breaking and the colliding particles as a thermal system, then it may be viewed as a phase transition due to a critical point in the phase diagram. This symmetry breaking implies in a first-order phase transition taking place at some $T=T_1$ whose experimental consequence is the transition of a slow decaying to a fast increasing of $\sigma_{tot}(s)$ as $s$ increases. On the other hand, the saturation of the Froissart-Martin bound as $s\rightarrow\infty$ may imply in a second phase transition taming the increasing of $\sigma_{tot}(s)$. If it occurs, then it is reasonable to suppose a second-order phase transition taking place at some $T=T_2$ in accordance with the second-order phase transition in momentum space \cite{antoniou_2}.

As mentioned, the transition from a decaying to a rising total cross section may be viewed as a first-order transition. The question that naturally arises is if this transition can correctly represent singularities in the momentum space (and vice-versa). If there exists a homotopy (topological defect) between the energy and momentum space interacting logarithmically, then it will produce the Kosterlitz-Thouless phase transition \cite{kt}. The mathematical consequence is the natural connection of singularities (geometrical properties) in both spaces. Hence, the fractal dimension observed in momentum space in \cite{antoniou_2} or in \cite{bialas,bialas_10} is correctly represented as a fractal dimension in $s$-space. Therefore, odderons and pomerons may be viewed as topological defects whose experimental evidence may be observed by intermittency patterns in momentum space \cite{antoniou_2}.

Therefore, the total cross section may be viewed as an energy-dependent multifractal structure. From a physical point of view, at low-energies the leading exchange process differs particle from antiparticle, since the emptiness measure in the collision process are distinct. Hence, the total cross section may be related to a measure of the emptiness in the elastic scattering. As the energy increases, the emptiness of the fractal structure vanishes and, therefore, $pp$ and $\bar{p}p$ total cross sections reach a minimum value. From this minimum, $pp$ and $\bar{p}p$ total cross sections start to be filled up and at high-energies tend to the same value, i.e., the fractal dimensions of both collision processes saturates the Froissart-Martin bound.



\section{\label{fr}Final Remarks}
The main goal of this paper is to analyze the multifractality of $pp$ and $\bar{p}p$ total cross sections. At low energies, $\sqrt{s}=2.5\sim20.0$ GeV,  the fractal dimension is negative and it may be viewed as a measure of the emptiness of the collision process (absence of a well-defined internal structure or less opaque in the impact parameter space). Above this range, as the energy $s$ increases, the fractal dimension turns positive, representing the filling up of $pp$ and $p\bar{p}$ scattering (presence of a well-defined internal structure or more opaque). Therefore, the collision particles appear to increase their matter density, with an approach to the black disk limit. 

In terms of Regge theory, the reggeon is the pole in the partial wave in $t$-channel of the scattering process. Here the negative fractal dimension may be associated to the odderon exchange and the positive value to the pomeron exchange. The odderon possesses negative parity and this will result in the observed differences between $pp$ and $\bar{p}p$ total cross sections at low energies. This result may be explained as follows. Considering the range $\sqrt{s}=2.5\sim 20.0$ GeV, the emptiness of $pp$ and $\bar{p}p$ scatterings decrease, i.e., the local fractal dimension approaches zero (the physical transition take place at $s_{min}$) and the structure becomes more opaque (well-organized). The decreasing of the total cross section in this range is, therefore, an increasing of information about the total cross section (from a less to a more detailed structure). Above $s_{min}$, the fractal dimension becomes positive, causing the amount of information about the total cross section to increase (a more well-organized structure). 

It is important to stress that fractal structure presented here is in according with the results of \cite{bialas} and \cite{antoniou_1,antoniou_2} obtained in momentum space. Therefore, the fractal dimension is not merely a point of view but reflects a global characteristic of the scattering. The Kosterlitz-Thouless phase transition may be useful to understand how the topological defects of the momentum space can be connected to the energy space (total cross section). If there is such connection, then the odderon/pomeron entity may be viewed as a topological defect whose experimental evidence can be pursued in the intermittency phenomenon (particle production). This phenomenon possesses a power-law control in the momentum space and it may imply in the same effect in the energy space. Therefore, a power-law may emerge as a mechanism to tame the fast increase of the total cross section as $s\rightarrow\infty$.

In the point of minimum, small deviations $\Delta s$ will not change $\sigma_{tot}(s)$. Therefore, the transition odderon/pomeron may be characterized by a flat fractal dimension, $b(\bar{b})\rightarrow 0$. The meaning of that is simple: the energy starts to be concentrated in a more well-defined region in the $s$-space. Furthermore, the transition from the negative to a positive fractal dimension may be a reflex of a phase-transition in the momentum space whose experimental effect is the rise of the total cross section.

Again, it is important to remember that the recent result to $pp$ total cross section obtained by the TOTEM collaboration may lead to the violation of the Froissart-Martin bound. However, one argues that this violation is a local phenomenon. Adopting asymptotic parametrizations procedures one neglects the physics of scattering processes at intermediate energies, i.e., when $s\rightarrow\infty$, it makes the small (or not so small) perturbations of the system to become flat, avoiding the access to phase transitions, for instance. Therefore, fast rises of the total cross section at intermediate energies may lead to the violation of asymptotic bounds and theorems. However, these formal results must be satisfied at $s\rightarrow\infty$ for internal consistency of Axiomatic Quantum Field Theory.

The physical explanation of the total cross section at low and intermediate energies is an open problem, despite the recent experimental results. However, even at high energies the scenario is not so different. The increase of total cross section as the energy increases represents a logarithm divergence. Therefore,there must exist some unknown physical mechanism in order to tame this rising. We are presently investigating this subject from the thermodynamical point of view.

\section*{Acknowledgement}
FSB thanks to CNPq for financial support and SDC thanks to UFSCar and to V.A. Okorokov for the first reading of this manuscript an useful comments.

\end{document}